\documentstyle[12pt,a4wide]{article}
\begin{document}
\renewcommand{\thefootnote}{\fnsymbol{footnote}}
\begin{flushright}
Imperial/TP/93-94/09 \\
hep-ph/9412217 \\
$2^{nd}$ December 1994
\end{flushright}

\vskip 1cm
\begin{center}
{\Large\bf A Re-examination of the Path Ordered Approach to Real Time
Thermal Field Theory}
\vskip 1.2cm
{\large\bf T.S.Evans\footnote{E-Mail: T.Evans@IC.AC.UK} \&
A.C.Pearson\footnote{E-mail: A.Pearson@IC.AC.UK}}\\
Blackett Laboratory, Imperial College, Prince Consort Road,\\
London SW7 2BZ  U.K. \\
\end{center}

\vskip 1cm
\begin{abstract}
We argue that the asymptotic condition should not be applied in the
derivation of the real time formalism of thermal field theory. It is
shown that, contrary to popular belief, the generating functional of
time ordered Green functions does not factorise. When no asymptotic
condition is applied to the real time formalism, we find that the
normal two component Feynman rules arises naturally. In addition, the
extra Feynman rule that is applied when calculating vacuum diagrams is
simply derived. We also clear up any doubts regarding the equivalence
of the real and imaginary time formalisms. Finally, we consider these
results in the case of the new real time contour.
\end{abstract}

\vskip 1cm
\renewcommand{\thefootnote}{\arabic{footnote}}
\setcounter{footnote}{0}

\section{Introduction}

The real time approach to thermal field theory, derived using the
techniques of path integrals \cite{Rivers,LvW,Tanguy,NS}, allows time
ordered Green functions to be calculated directly.  When using the
imaginary time formalism \cite{Kapusta}, one must perform some form of
analytic continuation from to real energies. This operation is in
principle highly non-trivial because the analytic structure at finite
temperature is much more complicated than at zero temperature. The
real time thermal Feynman rules with their doubling of the degrees of
freedom have produced results that are repeatable using other
formalisms and it is widely held that  the real time formalism,
derived using path integrals, is equivalent to other methods such as
thermo field dynamics or the Matsubara formalism.

However, there are many areas of doubt regarding the real time
formalism. The standard derivation of the real time Feynman rules
relies upon our being able to ignore certain sections of the contour
associated with the real time formalism. It is unclear as to whether
we are affecting the results of calculations by ignoring these
sections. Another worrying aspect is the application of the so-called
asymptotic condition to the generating functional of real time thermal
Green functions. No such condition is made upon other formalisms (such
as the imaginary time formalism). As well as this, we must apply
special Feynman rules when calculating vacuum diagrams\cite{TSEZPC}.
There is no natural explanation for these extra Feynman rules within
the standard formulation of the real time formalism. In fact, to
derive the extra rules in \cite{TSEZPC}, the author had to assume that
the external source terms in generating functional were, for vacuum
diagrams, time independent. This assumption is impossible to reconcile
with the time dependant asymptotic condition.

Finally, a recent paper by Xu \cite{Xu} reported some problems
associated with the standard path ordered approach to real time
thermal field thory. It was claimed that the imaginary and real time
formalisms were not equivalent. This is surprising since the two
formalisms differ only in the choice of curve applied to the
generating functional and so are encoding the same physics in
different ways.

We therefore think that the standard derivation of real time thermal
field theory is in need of re-examination. In this paper, we argue
that it is the asymptotic condition that is at the root of these
problems. We show by contradiction that the generating functional does
not factorise (the main reason for applying the asymptotic condition).

We then analyse the real time formalism without applying the
asymptotic condition. We find that the standard two component Feynman
rules arise naturally without any recourse to factorisation. We are
also able to show the reason for the extra Feynman rules that apply to
vacuum diagrams.  Once it is realised that the generating functional
does not factorise, a simple explanation also arises for the results
of  \cite{Xu}.

Finally, we analyse the real time formalism using the new time contour
\cite{TSENRTF}. We find that the standard two
component formalism also arises naturally out of this formalism.
However, we point out that the use of the $\epsilon$-condition in
this formalism causes this formalism to break down when calculating
vacuum diagrams.

\section{The Real Time Formalism}

In order to explain fully the apparent inconsistencies in the
derivation of the real time Feynman rules, we feel it is necessary to
briefly review the main points. This will also help to clarify the
notation used. While we hope to discuss all of the salient points, a
full discussion can be obtained in Landsman and van Weert \cite{LvW} or
Furnstahl and Serot \cite{FS}. Our discussion shall be in terms of a
scalar field theory. However, the main results apply to other fields
as well. Our starting point shall be the generating functional of path
ordered thermal Green functions.
\begin{eqnarray}
Z[J]&=
&\exp\bigg{(} \int_{C} \; dt V[-\imath\frac{\delta}{\delta J}]
\bigg{)}.Z_{0}[J]
\nonumber \\
Z_{0}[J] &=
&Z_{0}[0]\exp \bigg{\{}-\frac{\imath}{2}\int_{C}\! dt dt' J(t)
\Delta_{C}(t-t') J(t') \bigg{\}} \label{genfun}
\end{eqnarray}
Here $V[\phi]$ is the interaction potential and $\Delta_{C}$ is the
free field thermal propagator which satisfies the KMS
condition \cite{KMS}. We have suppressed the spatial indices for
notational convenience. The contour $C$ in the complex time plane is
arbitrary except for two conditions.
\begin{enumerate}
\item The starting point $t_{i}$ and the end point $t_{f}$ are related
by $t_{f}=t_{i}-\imath\beta$.
\item In order that the Green functions are bounded as their
time arguments are taken to infinity,
the curve must have a monotonically decreasing
or constant imaginary part.
\end{enumerate}
Green functions, $G_{n}$, are generated by functional differentiation
of $Z[J]$ with respect to the external source terms $J({\bf x},t)$.
\begin{equation}
G_{n}(t_{1},.....,t_{n})= \mbox{Tr}\bigg{(}  T_{c}
[\phi(t_{1})....\phi(t_{n}) ] e^{-\beta H} \bigg{)} =
\frac{1}{Z[0]}(-\imath)^{n} \frac{\delta^{n} Z[J]}{\delta
J(t_{1})....\delta J(t_{n})} \bigg{|}_{J=0}
\end{equation}
$T_{C}$ denotes path ordering. That is the fields are ordered with
respect to their positions along the contour $C$. This is the contour
equivalent of the usual time ordering.

It should be noted that the Green functions will in general depend
upon the contour chosen. This can lead to apparent differences between
different formalisms \cite{TSEGF} of thermal field theory. However,
this difference is due to the fact that the Green functions of
different formalisms are time ordered with respect to different
contours. Once we specify precisely what we are calculating, the result
will be independent of the contour chosen. We should therefore choose
the contour so as to suit the problem at hand. The contours associated
with the real-time formalism ($C_1 \oplus C_2 \oplus C_3 \oplus C_4$)
and imaginary-time (or Matsubara) formalism ($C_I$) are shown in
fig.\ref{fig:RTF}.
\begin{figure}[htb]
\setlength{\unitlength}{0.01in}%
\begin{picture}(490,360)(-30,0)
\thicklines
\put( 80,80){\circle*{10}}
\put( 80,270){\circle*{10}}
\put(480,270){\circle*{10}}
\put(480,80){\circle*{10}}
\put( 80,20){\circle*{10}}
\put( 40,280){\vector( 1, 0){480}}
\put( 80,270){\vector( 1, 0){280}}
\put(480,270){\vector( 0,-1){ 70}}
\put(480,80){\vector(-1, 0){280}}
\put( 80,80){\vector( 0,-1){ 40}}
\put(360,270){\line( 1, 0){120}}
\put(480,200){\line( 0,-1){120}}
\put(200,80){\line(-1, 0){120}}
\put( 80,40){\line( 0,-1){ 20}}
\put(480,280){\line( 0, 1){ 15}}
\put( 80,280){\line( 0, 1){ 15}}
\put(280,20){\line( 1, 0){ 15}}
\put(280,0){\vector( 0, 1){360}}
\put(345,245){\makebox(0,0)[lb]{\Large $C_1$}}
\put( 50,50){\makebox(0,0)[lb]{\Large $C_4$}}
\put(285,285){\makebox(0,0)[lb]{\Large $0$}}
\put(285,90){\makebox(0,0)[lb]{$-\imath(1-\alpha)\beta$}}
\put( 60,305){\makebox(0,0)[lb]{\Large $-{\cal T}/2$}}
\put(460,305){\makebox(0,0)[lb]{\Large ${\cal T}/2$}}
\put(305,10){\makebox(0,0)[lb]{\Large $ - \imath \beta$}}
\put(205,95){\makebox(0,0)[lb]{\Large $C_2$}}
\put(445,200){\makebox(0,0)[lb]{\Large $C_3$}}
\put(500,250){\makebox(0,0)[lb]{\Large $\Re e (\tau)$}}
\put(295,330){\makebox(0,0)[lb]{\Large $\Im m (\tau)$}}
\multiput(285,85)(-0.40000,-0.40000){26}{\makebox(0.4444,0.6667){\sevrm .}}
\multiput(275,85)(0.40000,-0.40000){26}{\makebox(0.4444,0.6667){\sevrm .}}
\put(240,20){\circle{10}}
\put(240,270){\circle{10}}
\put(245,160){\makebox(0,0)[lb]{\Large $C_{I}$}}
\put(240,160){\line( 0,-1){140}}
\put(240,270){\vector( 0,-1){110}}
\end{picture}
\caption{The time contours associated with the real and imaginary time
formalisms.}
\label{fig:RTF}
\end{figure}

\subsection{The thermal propagator}
We will now discuss the behaviours of the propagator at non-zero
temperatures. For our example of a scalar field, the propagator
satisfies
\begin{equation}
\bigg{(} \Box_c + m^{2} \bigg{)} \Delta_{C}({\bf
x-x'},t-t')=-\delta({\bf x-x'})\delta_{C}(t-t')
\label{wave}
\end{equation}
subject to the well known KMS boundary condition \cite{KMS}. Time
derivatives are defined along the contour $C$. $\delta_{C}(t-t')$ is
the generalisation of the normal Dirac delta function and satisfies $
\int_{C}\! dt \delta_{C}(t-t')=1$. To solve Eqn.\ref{wave}, we use the
following ansatz.
\begin{equation}
\Delta_{C}(t-t')=\Delta^{>}(t-t') \Theta_{C}(t-t') + \Delta^{<}(t-t')
\Theta_{C}(t'-t)
\end{equation}
 The function $\Theta_{C}$ is the contour version of the normal
step function and is defined by
\begin{equation}
\Theta_{C}(t-t')=\int_{C}^{t}\! dt'' \; \delta_{C}(t''-t')
\end{equation}
The KMS condition\cite{KMS} may be written as
\begin{equation}
\Delta^{>}(t-t')=\Delta^{<}(t-\{t'-\imath\beta\})
\end{equation}
Solving Eqn.\ref{wave},we find that for our scalar field theory the
free field propagator expressed in terms of time and three-momentum is
given by
\begin{eqnarray}
\imath\Delta(t-t',{\bf k}) &=&
-\frac{\imath}{2\omega}\frac{1}{1-e^{-\beta\omega}} \bigg{[}\bigg{(}
e^{-\imath\omega (t-t')}+ e^{-\beta\omega}e^{\imath\omega
(t-t')}\bigg{)}\Theta_{C}(t-t') \nonumber \\
&&+ \bigg{(} e^{\imath\omega (t-t')} + e^{-\beta\omega} e^{-\imath\omega
(t-t')} \bigg{)} \Theta_{C}(t'-t) \bigg{]} \label{solution}
\end{eqnarray}
where $\omega^{2}={\bf k}^{2}+m^2$. This can also be represented in
its spectral form \cite{Mills}.
\begin{eqnarray}
\imath\Delta_{C}(t-t',{\bf k}) &=&
\int \frac{dk_{0}}{2\pi} \rho(k_{0},{\bf k}) e^{-\imath k_{0} (t-t')}
\big{[} \Theta_{C}(t-t')+N(k_{0}) \big{]} \label{spec}\\
\rho(k_{0},{\bf k}) &=& 2\pi sgn(k_{0}) \delta(k_{0}^{2}-\omega^{2})
\nonumber
\end{eqnarray}
$N(k_{0}$) is the Bose-Einstein distribution and
$sgn(k_{0})=\frac{k_{0}}{|k_{0}|}$.

\subsection{The asymptotic condition}
In order to derive the Feynman rules, we must first consider the
notion of factorisation. We would like to see whether the
contribution to the generating functional from the sections $C_{3}$
and $C_{4}$ can be separated from the contribution from $C_{1}$ and
$C_{2}$ (see fig.\ref{fig:RTF}). That is whether we can write the
generating functional as
\begin{equation}
Z[J]=Z_{12}[J].Z_{34}[J]
\end{equation}
where in $Z_{ab}[J]$, the time integrations are constrained to lie
along $C_{a}\oplus C_{b}$. We would like to be able to do this since
if factorisation does occur, then only $Z_{12}[J]$ will contribute to
the time ordered Green functions. $Z_{34}[J]$ does not depend on any
real times and as such behaves as an overall normalisation factor
which is irrelevant to the calculation of Green functions.

Looking at Eqn.\ref{genfun}, we can see that the full interacting
generating functional will factorise if the free field case does.
Examining $Z_{0}[J]$ in closer detail we can see that factorisation
occurs if
\begin{equation}
\int_{C_{1,2}}\! dt\int_{C_{3,4}}\! dt' J(t)\Delta_{C}(t-t') J(t')
=0
\label{factcon}
\end{equation}
To see whether this result is true we must study the properties of
the propagator.

Using Eqn.\ref{solution}, it has been shown \cite{FS} that the
propagator tends to zero at large time differences. However despite
this fact, they also showed that certain products of propagators do
not vanish in this large time limit. This result is due to the fact
that the spectral density in Eqn.\ref{spec} is a generalised function.
As a result, the Feynman rules produced by this formalism are
unwieldy. To avoid this problem, the $\epsilon$-prescription is
introduced \cite{LvW,FS}.  We modify the spectral density in the
following way.
\begin{eqnarray}
\rho(k_{0},{\bf k}) &=&
2\pi sgn(k_{0}) \delta_{\epsilon}(k_{0}^{2}-\omega^{2}) \\
2\pi\delta_{\epsilon}(X) &=&
\frac{\imath}{X+\imath\epsilon}-\frac{\imath}{X-\imath\epsilon}
\end{eqnarray}
It is understood that we must hold $\epsilon$ finite throughout any
calculation and only take the limit $\epsilon\rightarrow 0$ at the end
of any calculation. The effect of this regularisation is to modify the
propagator.
\begin{equation}
\Delta_{C}(t-t',{\bf k})\rightarrow \Delta_{C}(t-t',{\bf k}).
e^{-\epsilon |t-t'|}
\end{equation}
We can see that we recover the correct form in the limit
$\epsilon\rightarrow 0$.

Using this modified propagator in Eqn.\ref{factcon}, we can see that
the only non-zero contribution comes from the region $Re(t)
\rightarrow \pm \infty$. If in addition we restrict the source terms
$J(t)$ to be zero in this limit then Eqn.\ref{factcon} will be
satisfied. This is the so-called asymptotic condition.
\begin{equation}
\lim_{Re(t)\rightarrow \pm\infty,\/t\in C_{1},C_{2}} J(t)=0
\label{ac}
\end{equation}

\subsection{Real Time Feynman rules}

To calculate real time, time ordered Green functions we need only
consider $Z_{12}[J]$.
\begin{equation}
Z_{12}[J]=\exp\bigg{\{}\imath\int_{C_{1,2}}dt
V[-\imath\frac{\delta}{\delta
J}]\bigg{\}}\exp\bigg{(}-\frac{\imath}{2}\int_{C_{1,2}} dtdt'
J(t)\Delta_{c}(t-t')J(t') \bigg{)}
\end{equation}
We can rewrite this purely in terms of real times.
\begin{equation}
Z_{12}[J]=\exp\bigg{\{} \imath\int_{-\infty}^{\infty} dt
V[-\imath\frac{\delta}{\delta J_{1}}] - V[-\imath\frac{\delta}{\delta
J_{2}}]\bigg{\}} \exp\bigg{(}
-\frac{\imath}{2}\int_{-\infty}^{\infty}dtdt'
J_{a}(t)\Delta_{ab}(t-t') J_{b}(t') \bigg{)}
\label{fr12}
\end{equation}
where the source term $J_{a}(t)$ and the propagator $\Delta_{ab}(t-t')$
are defined by
\begin{eqnarray}
J_{a}(t)&=&\left\{ \begin{array}{ll}
J(t) & a=1 \\
- J(t-\imath \{1-\alpha\}\beta) & a=2
\end{array}
\right. \\
\Delta_{ab}(t-t')&=&\Delta_{C}(\tau-\tau') \left\{ \begin{array}{ll}
\tau(')=t(') & a (b)=1 \\
\tau(')=t(')-\imath\{1-\alpha\}\beta & a(b)=2
\end{array}
\right.
\end{eqnarray}

We can now read off the real time thermal Feynman rules. We can see
that we have, in effect, doubled the degrees of freedom of our theory.
At each vertex we assign a thermal index (type $1$ or $2$
corresponding to the section $C_{1}$ and $C_{2}$ of the real time
contour). Type $1$ vertices correspond to real particles. Vertices
only couple fields of the same thermal index, the coupling differing
only in the relative sign between the two thermal types. The two types
of field only couple through the thermal propagator which now becomes
a two by two matrix.

Using our example of a scalar field theory (now with a cubic
interaction), the Feynman rules are
\begin{eqnarray}
\begin{picture}(40,50)(0,0)
\put(0,3){\line(1,0){20}}
\put(20,3){\line(3,2){20}}
\put(20,3){\line(3,-2){20}}
\put(15,5){$a$}
\end{picture}
&=& (-1)^{a}\imath\lambda \\
\begin{picture}(50,50)(0,0)
\put(7,3){\line(1,0){40}}
\put(0,0){$a$}
\put(48,0){$b$}
\end{picture}
&=& \imath\Delta_{ab}(k)
\end{eqnarray}
\begin{eqnarray}
\imath\Delta_{11}(k)&=&\frac{\imath}{k^{2}-m^{2}+\imath\epsilon} + 2\pi
N(|k_{0}|) \delta_{\epsilon}(k^{2}-m^{2}) \label{prop11}\\
&&\nonumber\\
\imath\Delta_{12}(k)&=& \frac{\pi}{\omega} e^{\frac{\beta}{2}|k_{0}|}
N(|k_{0}|)
\bigg{[} e^{(\alpha-\frac{1}{2})\beta |k_{0}|}
\delta_{\epsilon}(k_{0}-\omega) + e^{-(\alpha-\frac{1}{2})\beta
|k_{0}|} \delta_{\epsilon}(k_{0}+\omega) \bigg{]} \\
&&\nonumber\\
\imath\Delta_{21}(k)&=&
\frac{\pi}{\omega} e^{\frac{\beta}{2}|k_{0}|} N(|k_{0}|)
\bigg{[} e^{-(\alpha-\frac{1}{2})\beta |k_{0}|}
\delta_{\epsilon}(k_{0}-\omega) + e^{(\alpha-\frac{1}{2})\beta
|k_{0}|} \delta_{\epsilon}(k_{0}+\omega) \bigg{]} \\
&&\nonumber\\
\imath\Delta_{22}(k)&=&\frac{-\imath}{k^{2}-m^{2}-\imath\epsilon} + 2\pi
N(|k_{0}|) \delta_{\epsilon}(k^{2}-m^{2})
\end{eqnarray}
The choice $\alpha=\frac{1}{2}$ is normally used as for this choice
the propagators take on a particularly simple form.

It should be noted that we have used $|k_{0}|$ in the argument of the
terms proportional to the $\delta_{\epsilon}$ functions, and not
$\omega$. This is a result of the use of the $\epsilon$-prescription
\cite{FS}. If we had set $\epsilon$ to zero, we would have a Dirac
delta function in Eqn.\ref{prop11} instead of the regularised
$\delta_{\epsilon}$ function. The term proportional to this Dirac
delta functional would then only be defined on mass-shell.  In that
case there is no difference between using $\omega$ or $|k_{0}|$ in the
argument of the Bose-Einstein function. However, since we have kept
$\epsilon$ finite, this term must be analytically continued off
mass-shell so as to be consistent with the KMS condition \cite{KMS}.
For this reason, $|k_{0}|$ and not $\omega$ must be used. It is not,
as is stated in \cite{Niegawa}, a scheme for ignoring the sections
$C_{3}$ and $C_{4}$ of the real time contour.

\section{Does $Z[J]$ Factorise?}
Despite the successes in applying the real time thermal Feynman rules
to physical problems, there are a number of problems raised by the use
of the asymptotic condition. No such condition is made upon the
generating functional used in the imaginary time formalism.  It is not
unreasonable to question whether forcing the asymptotic condition upon
the real time formalism changes the results derived using it. If we
could show that the generating functional is indeed changed by using
the asymptotic condition then we would be forced to remove this
condition and with it the notion of factorisation.

To proceed, let us assume that we can use the asymptotic condition
and that the generating functional factorises.
\begin{equation}
Z[J]=Z_{12}[J]Z_{34}[J] \label{factor1}
\end{equation}
We now note that since the sections $C_{3}$ and $C_{4}$ of the real
time contour (see fig.\ref{fig:RTF}) are separated by an infinite
time, we may further factorise the contributions to $Z_{34}[J]$ from these two
sections. That is the generating functional may be factorised
further.
\begin{equation}
Z[J]=Z_{12}[J]Z_{3}[J]Z_{4}[J] \label{factor}
\end{equation}
This is a remarkable result. We know that $Z_{12}[J]$ is independent
of the parameter $\alpha$ since it has been shown that the Green
functions derived from the $1$-$2$ sector do not depend on $\alpha$
 \cite{LvW,FS}. This would then imply that the product
$Z_{3}[J]Z_{4}[J]$ is also $\alpha$ independent. If in fact
$Z_{3}[J]Z_{4}[J]$ is $\alpha$ dependant then we have shown by
contradiction that the generating functional of real time thermal
green functions cannot be factorised.

To test the $\alpha$ dependance of $Z_{3}[J]Z_{4}[J]$ let us set the
external sources to zero in Eqn.\ref{factor}. Then, making use of the
result $Z_{12}[J=0]=1$ \cite{LvW,FS} and the fact that $Z$ is
independent of the precise curve $C$ chosen, we have that
\begin{equation}
Z[0]=Z_{ITF}[0]=Z_{3}[0]Z_{4}[0] \label{wrong}
\end{equation}

The interpretation of Eqn.\ref{wrong} is that if we calculate any
vacuum diagram in the imaginary time formalism, we can repeat the
same result by adding together the contributions of the same diagram
calculated separately on the sections $C_{3}$ and $C_{4}$ of the
real time contour. It should be noted that we implicitly assume that
the $\epsilon$-prescription is applied to both the imaginary and
real time calculations. However since the time differences involved
are finite the introduction of the $\epsilon$-prescription
will not affect the results.

As an example we shall use the simple case of a scalar field with a
cubic interaction. One such vacuum diagram is

\begin{equation}
\setlength{\unitlength}{0.5pt}
\begin{picture}(200,50)(0,0)
\thicklines
\put(60,5){\circle{40}}
\put(160,5){\circle{40}}
\put(80,5){\line(1,0){60}}
\put(85,10){$t$}
\put(130,10){$t'$}
\end{picture}
=V=
-\frac{\lambda^{2}}{8}\bigg{[} \int \frac{d^{3}\! k}
{(2\pi)^{3}} \imath\Delta(t-t'=0,{\bf k}) \bigg{]}^{2} \int_{C} dtdt'
\imath\Delta(t-t',{\bf k=0})
\end{equation}
Evaluating this integral using the imaginary time formalism we find
\begin{equation}
V_{ITF}=\frac{\lambda^{2}}{8}\bigg{[} \int\frac{d^{3}\! k}{(2\pi)^{3}}
\bigg{(} \frac{1}{2\omega}+\frac{n(\omega)}{\omega}
\bigg{)}\bigg{]}^{2} \frac{\beta}{m^{2}}
\end{equation}
And the sum of the contributions from $C_{3}$ and $C_{4}$ is
\begin{equation}
V_{3}+V_{4}=\frac{\lambda^{2}}{8}\bigg{[} \int\frac{d^{3}\!
k}{(2\pi)^{3}} \bigg{(} \frac{1}{2\omega}+\frac{n(\omega)}{\omega}
\bigg{)}\bigg{]}^{2}
\bigg{\{}
\frac{\beta}{m^{2}}-\frac{2}{m^{3}}\frac{(1-e^{-\alpha\beta
m})(1-e^{-(1-\alpha)\beta m})}{(1-e^{-\beta m})}
\bigg{\}}
\end{equation}

Clearly these results differ. The contribution from the sections
$C_{3}$ and $C_{4}$ of the real time contour is $\alpha$ dependant
contrary to Eqn.\ref{factor}. Note however, that for $\alpha=0,1$
the two results do concur. This is trivially so since for these
values of $\alpha$, either $C_{3}$ or $C_{4}$ becomes the Matsubara
contour. For these two values of $\alpha$ the generating functional
does factorise. However, factorisation in these two cases does not
rely on the asymptotic condition but on the boundary condition
$\phi(t)=\phi(t-\imath\beta)$ imposed on the field configurations in
the path integral.

This result shows by contradiction that the generating functional
does not factorise. If we did not apply the asymptotic condition then
there would be extra contributions to this diagram coming from the
case where one time belongs to $C_{1}\oplus C_{2}$ and the other
belongs to $C_{3}\oplus C_{4}$. These extra contributions would
restore the equivalence between the imaginary and real time
formalisms\cite{APBanff}.

\section{The RTF without the asymptotic condition}
The fact that the generating functional does not factorise raises a
number of points. The real time Feynman rules have been successfully
used for a number of years and gives results that are repeatable using
other formalisms. We must assume that in almost all cases, real time
thermal Feynman rules, based on a doubling of the degrees of freedom,
are correct. It must therefore be possible to derive the two component
Feynman rules from the full, unfactorised generating functional
without any recourse to the asymptotic condition. We should also be
able to discover whether these Feynman rules break down in certain
types of calculations. If we are able to find situations where these
rules break down, then this raises questions about the applicability
of other real time formalisms, such as Thermo Field
Dynamics\cite{Rivers,LvW}, to these problems. Thermo field dynamics
produces the same Feynman rules as the real time formalism and as such
will also give incorrect results in these areas where the path ordered
real time approach breaks down.

Let us now analyse the real time formalism without making use of the
asymptotic condition. Since we cannot assume that the generating
functional factorises, we must account for all four sections of the
real time contour. In principle we must deal with not a two component
theory but a four component formalism! However as we shall see, great
simplifications can be made for time ordered Green functions. As a
simple example let us consider one contribution to the two point Green
function of our scalar field theory, represented by the diagram
\begin{equation}
\begin{picture}(50,30)(0,0)
\put(5,3){\line(1,0){10}}
\put(25,3){\circle{20}}
\put(35,3){\line(1,0){10}}
\put(0,0){t}
\put(46,0){t'}
\end{picture}
= -\frac{\lambda^{2}}{2}\int_{C}\!d\tau\int_{C}\!d\tau '
\Delta(t-\tau)\Delta^{2}(\tau-\tau ') \Delta(\tau '-t')
\label{bubble}
\end{equation}
Since we cannot now assume that we need only consider the contribution
from the $1$-$2$ sector of the real time contour, the time integrals
in Eqn.\ref{bubble} will be over all four sections of the curve.
However, if we assume that one of the external times is real and
finite then we may safely ignore the contributions to this integral
from the sections $C_{3}$ and $C_{4}$. This is because the
$\epsilon$-regularisation damps out any contributions over large time
scales. We can therefore see that the contribution to this diagram
from the full real time contour comes only from the $1$-$2$ sector.

We can generalise this argument to an n-point Green function. By
using the Schwinger-Dyson equation, we can extract vertices from
within the Green function. For example in our example of a scalar
field coupled by a cubic self interaction, the Schwinger-Dyson
equation for the two-point Green function is represented pictorially by
\begin{eqnarray}
\begin{picture}(40,30)(0,0)
\put(0,3){\line(1,0){10}}
\put(17,3){\circle*{15}}
\put(25,3){\line(1,0){10}}
\put(-5,0){t}
\put(36,0){t'}
\end{picture}
&=
&
\begin{picture}(45,30)(0,0)
\put(0,3){\line(1,0){35}}
\put(-5,0){t}
\put(36,0){t'}
\end{picture}
+
\begin{picture}(70,30)(0,0)
\put(5,3){\line(1,0){30}}
\put(42,3){\circle*{15}}
\put(50,3){\line(1,0){10}}
\put(25,3){\line(0,1){10}}
\put(25,20){\circle*{15}}
\put(0,0){t}
\put(22,-5){$\tau$}
\put(61,0){t'}
\end{picture}
+
\begin{picture}(70,30)(0,0)
\put(5,3){\line(1,0){20}}
\put(25,3){\line(5,2){10}}
\put(25,3){\line(5,-2){10}}
\put(40,3){\circle*{15}}
\put(47,3){\line(1,0){10}}
\put(0,0){t}
\put(58,0){t'}
\put(22,-5){$\tau$}
\end{picture}
\nonumber\\
&&\nonumber\\
G_{2}(t,t')&=& \imath\Delta(t-t') + \lambda \int_{C} d\tau
\Delta(t-\tau)\bigg{(} G_{1}(\tau)
G_{2}(\tau,t')+\frac{1}{2}G_{3}(\tau,\tau,t') \bigg{)} \label{only12}
\end{eqnarray}
We have suppressed spatial indices for notational convenience.
$G_{n}(t_{1},...,t_{n})$ is the n-point Green function.

If we fix the external time $t$ to be real and finite then although
the $\tau$ integral in Eqn.\ref{only12} is over the entire real time
contour $C$, only the sections $C_{1}$ and $C_{2}$ will contribute to
this integral. The sections $C_{3}$ and $C_{4}$ are infinitely far
away and as such we may ignore them due to the
$\epsilon$-prescription. The $\epsilon$-prescription therefore allows
us to restrict $\tau$ to lie on the sections $C_{1}$ and $C_{2}$.

We can repeat this method to extract more vertices from the two-point
Green function. The same argument can be used to fix these vertices to
lie on $C_{1}$ or $C_{2}$. In principle, this argument is infinitely
recursive. However to a given order in perturbation theory this
process has a finite number of steps. The final result is that if we
fix one external leg to finite times then we fix {\em every} internal
and external time to belong to either $C_{1}$ or $C_{2}$.

Armed with this result we can now make statements about the range of
validity of the standard real time formalism. The real time formalism
should break down dealing with diagrams where none of the external
times are fixed to be finite. This means that real time time-ordered
Green functions can be correctly calculated using the normal
two-component Feynman rules. However, as we have seen in the previous
section, this formalism will not calculate vacuum diagrams correctly.
Since there are no external legs, the internal time integrations must
be over the entire real time contour. The $1$-$2$ sector will be
insufficient.

It is already known that special techniques must be applied to
correctly evaluate vacuum diagrams using the real time formalism
\cite{LvW,TSEZPC,Um}. This extra Feynman rules involves fixing one of
the internal vertices to be type one. The other vertices of the
diagram will only then contribute in the $1$-$2$ sector because of the
arguments shown above. The last time integration (the vertex we have
fixed) is easily done since it is assumed that the argument of the
integration will be time independent because of thermodynamic
equilibrium. This last integration will just give an overall factor of
$-\imath\beta$ since the integration is over the entire real time
contour. It should be noted that in the derivation of this result in
\cite{TSEZPC} it was assumed that the external sources of the generating
functional were time independent This assumption is contrary to the
asymptotic condition and provides extra evidence that the asymptotic
condition  should be disregarded.

As a final point we should like to discuss the work of Xu \cite{Xu}
in which some problems associated with connected Green functions as
calculated within the standard path-ordered approach to real time
thermal field theory were reported. The basis of the discussion
hinged upon the fact that the generating functional of the imaginary
time formalism was not unitarily equivalent to the generating
functional, $Z_{12}[J]$, derived using only the sections $C_{1}$ and
$C_{2}$ of the real time contour, and as such the two formalisms were
not equivalent.

The work presented here shows that we may derive the standard real
time thermal Feynman rules from the full generating functional of the
real time formalism. We do not need to factorise $Z[J]$ and use only
the section $Z_{12}[J]$. Since the real and imaginary time formalisms
may be connected by an analytic continuation of the time contour used
we can see that the two formalisms are equivalent, contrary to the
result of Xu \cite{Xu}. If anything, our work suggests that it is
thermo field dynamics which is inequivalent to path ordered approaches
to thermal field theory, though only in special cases.

\section{The other real time contour}
Recently, a new contour has been proposed to derive the real time
formalism of thermal field theory. As can be seen in
fig.\ref{fig:NRTF}, the contour has only two sections. Instead of
two horizontal and two vertical sections, we now have one section
along the real axis and a second that has an infinitesimal gradient
so that the endpoint of the contour is a distance $-\imath\beta$
below the starting point as required.
\begin{figure}
\setlength{\unitlength}{0.01in}%
\begin{picture}(480,325)(40,315)
\thicklines
\put(160,340){\circle*{10}}
\put(160,530){\circle*{10}}
\put(540,530){\circle*{10}}
\put(320,320){\vector( 0, 1){270}}
\put(120,520){\vector( 1, 0){450}}
\put(160,530){\vector( 1, 0){240}}
\put(400,530){\line( 1, 0){140}}
\put(540,530){\vector(-2,-1){180}}
\put(360,440){\vector(-2,-1){200}}
\put(330,410){\line(-1, 1){ 20}}
\put(190,435){\makebox(0,0)[lb]{\Large $-i(1-\alpha)\beta$}}
\put(135,315){\makebox(0,0)[lb]{\Large $-\alpha {\cal T} - i \beta$}}
\put(135,545){\makebox(0,0)[lb]{\Large $-\alpha {\cal T}$}}
\put(500,545){\makebox(0,0)[lb]{\Large $(1-\alpha){\cal T}$}}
\put(305,500){\makebox(0,0)[lb]{\Large $0$}}
\put(525,495){\makebox(0,0)[lb]{\Large $\Re e (\tau)$}}
\put(335,590){\makebox(0,0)[lb]{\Large $\Im m (\tau)$}}
\put(390,540){\makebox(0,0)[lb]{\Large $N_{1}$}}
\put(370,425){\makebox(0,0)[lb]{\Large $N_{2}$}}
\end{picture}
\caption{The new real time contour (${\cal T}\rightarrow\infty $).}
\label{fig:NRTF}
\end{figure}
Since this contour has only two sections, the two component Feynman
rules naturally drop out without any use of the asymptotic condition
or of factorisation. However since we know that these Feynman rules
break down in certain circumstances and we also know that this new
contour should be equivalent to any other path ordered approach to
thermal field theory, we must examine this approach to the real time
formalism in more depth.

We proceed in a similar manner to the conventional real time
formalism. The generating functional may be manipulated in order to
write it in terms of a two component theory.
\begin{eqnarray}
Z[J] &= & \exp \bigg{(} \imath \int d^{4}x \sum_{a=1}^{2} (-1)^{a+1}
V_{a}[-\imath\frac{\partial}{\partial J_{a}(x)}]\bigg{)}. Z_{0}[J]
\nonumber\\
Z_{0}[J] &= & \exp \bigg{\{} \sum_{a,b=1}^{2}\int d^{4}x \int
d^{4}x' J_{a}(x) \Delta_{ab}(x-x') J_{b}(x') \bigg{\}}
\end{eqnarray}
where $J_{a}(x)$ and $\Delta_{ab}(x-x')$ are defined in the following
way.
\begin{eqnarray}
J_{a}(x) &= &\left\{ \begin{array}{ll}
J({\bf x},t) & a=1 \\
- J({\bf x},\tau) & a=2 \\
\end{array}
\right. \\
\Delta_{ab}(x-x') &= &\left( \begin{array}{cc}
\Delta({\bf x-x'},t-t') & \Delta({\bf x-x'}, t-\tau') \\
\Delta({\bf x-x'},\tau-t') & \Delta({\bf x-x'}, \tau-\tau')
\end{array}
\right) \label{NRTFprop}
\end{eqnarray}
We have used $\tau$ and $\tau'$ to denote that the time is along the
section $N_{2}$ of the new real time contour. We can parameterise $\tau$
and $\tau'$ in terms of real time parameters $t$ and $t'$ by
\begin{equation}
\tau=\bigg{(} 1 + \frac{\imath\beta}{{\cal T}} \bigg{)}t
-\imath(1-\alpha)\beta
\end{equation}
$\Delta({\bf x},t)$ is defined in equation \ref{solution}. Just as with the
conventional derivation of the real time formalism we have made use of
the $\epsilon$-prescription.

In the previous section we found that if we fixed the external times
of a given Green function then the conventional Feynman rules were
correct. In terms of the new contour this statement means that by
fixing the external times to be real, we may safely ignore the
infinitesimal gradient of the contour $N_{2}$. Examining the form of
the propagator in equation \ref{NRTFprop} we see that the terms
containing the gradient $\frac{\beta}{{\cal T}}$ will only contribute
to the result when large times are encountered. However, we make use
of the $\epsilon$-prescription to ignore the contributions from large
time scales. We can therefore see that by fixing the external times to
be finite the gradient of $N_{2}$ can be safely ignored. That is we
have again managed to show that the conventional two-component Feynman
rules apply in the calculation of any time ordered thermal Green
function. This result was arrived at separately from the conventional
real time thermal field theory and as such supports our earlier
results of section 3.

We now turn to the more problematic study of the calculation of vacuum
diagrams. In this case we cannot fix any of the times involved to be
real. We would therefore expect to have to account for the terms
containing the gradient of the contour $N_{2}$. However, as we have
stated earlier, these terms only contribute over large time scales
which is precisely where the $\epsilon$-prescription damps out any
effects.  This formalism is therefore {\em also} not able to correctly
evaluate vacuum diagrams directly, given that one uses the usual
Feynman rules i.e.\ if one takes ${\cal T}$ to infinity before one
takes $\epsilon$ to zero. The use of the $\epsilon$-prescription is
inconsistent with the gradient of $N_{2}$ in this case. Unlike the
conventional real time contour (see fig.\ref{fig:RTF}) the
$\epsilon$-prescription does indeed change the generating functional
derived using this new contour.

Of course, one can use the usual double field real time Feynman
rules to work out vacuum diagrams if we use a trick and calculate a
subset of the vacuum diagrams \cite{TSEZPC,Um}.

\section{Conclusions}

In this paper, we have attempted to make sense of the derivation of
the real time formalism and of the many claims regarding its
equivalence to other formalisms such as the Matsubara method and
thermo field dynamics. We have found that problems have arisen from
the erroneous use of the asymptotic condition. We have proved that the
real time generating functional does not factorise.  The main reason
for our desire to factorise the generating functional has been so that
we can express time ordered Green functions in terms of real time
arguments. However as we have shown in this paper, the standard
two-component Feynman rules arise from the full, unfactorised
generating functional. The asymptotic condition was used to force
factorisation upon the real time formalism and so guarantee the two
component formalism. As we have shown it is not just unnecessary to do
this but it also causes the real time generating functional to produce
incorrect results in certain cases. This has led to confusion about
the validity of real time techniques. Once the asymptotic condition is
dropped, we can clear up these confusing points. The full generating
functional must be equivalent to the generating functional of the
Matsubara Green functions. This is because the real time contour is
just an analytic continuation of the Matsubara contour. We have not
disregarded certain sections of the real time contour as is done in
the standard real time formalism and so our results are due to the
full real time contour.

The absence of factorisation has been shown to be precisely the result
we needed to clear up a number of points regarding the use of thermal
field theory. We have shown that the extra Feynman rules \cite{TSEZPC,Um}
needed to deal with vacuum diagrams arises naturally out of our
unfactorised formalism.

Finally, we have examined these results in terms of the new
real time contour \cite{TSENRTF}. We have again shown
that the real time thermal Feynman rules may be safely applied in the
calculation of time ordered thermal Green functions. However, we have
shown that the introduction of the $\epsilon$-prescription cancels out
any effect due to the infinitesimal slope of the section $N_{2}$ of
the new contour. As such, the new curve can only be used to evaluate
vacuum diagrams indirectly through use of the methods in
\cite{TSEZPC,Um}.

\section{Acknowledgements}
Andy would like to thank Tim Walker for being a bit skeptical about
factorisation.
We would also like to thank Chris van Weert and A. Niegawa for useful
discussions.

\end{document}